# Enhanced Stability of the Model Mini-protein in Amino Acid Ionic Liquids and Their Aqueous Solutions

Guillaume Chevrot,[1] Eudes Eterno Fileti,[2] and Vitaly V. Chaban[2]

1) MEMPHYS - Center for Biomembrane Physics, Department of Physics, Chemistry and Pharmacy, University of Southern Denmark, Campusvej 55, 5230 Odense, Denmark.

2) Instituto de Ciência e Tecnologia, Universidade Federal de São Paulo, 12231-280, São José dos Campos, SP, Brazil

**Abstract.** Using molecular dynamics simulations, the structure of model mini-protein was thoroughly characterized in the imidazolium-based amino acid ionic liquids and their aqueous solutions. We report that the mini-protein is more stable when AAIL is added as a cosolvent. Complete substitution of water by organic cations and anions further results in hindered conformational flexibility of the mini-protein. This observation suggests that AAILs are able to defend proteins from thermally induced denaturation. We show by means of radial distributions that the mini-protein is efficiently solvated by both solvents due to agood mutual miscibility. However, amino acid based anions prevail in the first coordination sphere of the mini-protein.

**TOC Graphic**

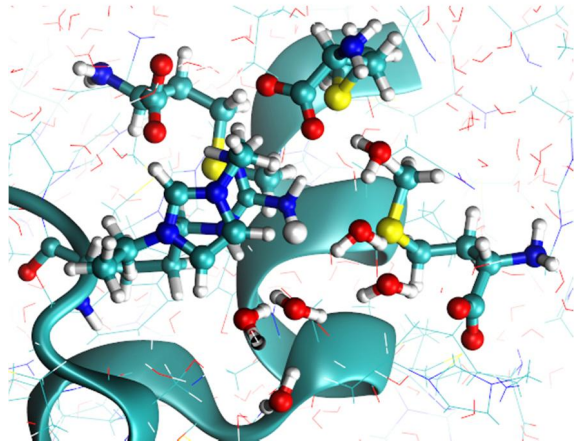

**Introduction**

Efficient extraction and preservation of proteins constitute important goals of modern biotechnology.[1-6] The extracted and preserved proteins must retain their structure and function, since many kinds of analysis are available in vitro only. This often appears impossible due to a high sensitivity of protein configurations to local intermolecular interactions. The three-dimensional structure of protein is maintained by means of hydrogen bonds and multiple hydrophobic interactions of slightly varying strength.[3,6] Electrostatic interactions involving polar amino acid residues and monovalent inorganic ions (chloride, sodium, potassium, etc) are also of high importance to stabilize proteins and provide acceptable level of hydration of the binding sites. Even tiny change in the environment may alter the fine balance of these interactions causing denaturation of protein.[7] Denaturation ultimately leads to unfolding and inactivation, but can also result in specific unnatural arrangements, whose function is either absent or far from the expected one. Many chronic diseases are linked to incorrect protein folding. The effect of such changes is close to mutations and may be fatal for the living organism.

An extensive evidence exists that solvation plays a critical role in stability and function of most proteins.[6-11] Accurate spatial information about proteins is obtained from routine diffraction experiments. Consistent interpretation of the diffraction density maps requires computer simulations with atomistic precision and adequate sampling of the phase space in the vicinity of the protein. In turn, the presence of macromolecular solute induces significant and permanent effects on microscopic structure and transport properties of the surrounded solvent and dissolved ions. Specific regions inside many proteins exist, where water cannot penetrate easily. The density of solvent in these regions is, consequently, different from the bulk density at given temperature and pressure. This feature is important to maintain the required configuration of an entire protein, as was found via atomistic molecular dynamics simulations.

Small fluctuations of local temperature and pressure can result in protein denaturation, since they perturb an accurate free energy balance. These same factors apply when a protein is extracted from the living organism and handled in vitro. Chemical modification, immobilization, additional stabilizing agents and genetic modification are major techniques to allow protein storage outside their natural environment.[3] The required temperature regime must be thoroughly maintained to prevent thermally induced structure alterations, most of which are strictly irreversible due to energy barriers.

Room-temperature ionic liquids (RTILs) comprise an organic cation and organic or inorganic anion.[12-15] In quite a few cases, these ionic combinations result in low melting points (below 373 K), which nevertheless comes at a price of a significantly high shear viscosity (ca. 100 cP). RTILs exhibit unique physical and chemical properties including negligible vapor pressure, non-flammability, high thermal and chemical stability, high ionic conductivity and interesting solvation behavior. RTILs are currently probed in versatile fields of chemical engineering, such as chemical sensors, electrochemical power sources, solution chemistry, gas capture, separation applications and catalysis.[16-20] Certain ionic liquids are environmentally friendly, without any clearly determined toxicity in relation to living cells. There exists experimental evidence that grafting of ether groups to RTILs decreases toxic effects. Successful applications of RTILs in biotechnology are known involving biological catalysis and protein conservation. It was argued that RTILs can serve as highly efficient participants in versatile biochemical processes, in addition to providing a well tunable reaction media. Water cannot be completely removed from the reaction environment, therefore researchers have to deal with biphasic systems: RTIL + water. According to recent reports, aqueous solutions of RTILs foster protein activity and protein stability. Some RTILs prevent or inhibit protein aggregation and improve refolding. An ability of ionic liquids to exclude aggregating of proteins (for instance, during folding) is an extremely important feature. Such a behavior implies strong specific interactions between proteins and RTILs, which may be sometimes more favorable than those between proteins and water. Interactions, observed in the protein solvated in RTIL and water, are not perfectly understood currently, whereas mostly speculative interpretations are available. Molecular dynamics simulations are able to provide significant assistance along these lines.

Following the conclusions of Herrman and coworkers,[21] solvent properties of solutions change from typical electrolyte solution-like behavior to molten salt-like behavior as a function of RTIL concentration. Phase transition can be hypothesized in this case. Bhattacharyya and coworkers[22] used femtosecond up-conversion to study solvation dynamics of a probe covalently attached via the lone cysteine group to a protein, in the presence of methylimidazolium RTIL. The average solvation time decreased from 650 ps to 260 ps, as 1.5 RTIL was supplied. The authors ascribed this observation to protein unfolding, which made the probe more exposed. Rama Devi and coworkers[23] concluded that stability of proteins and their functional groups can be drastically modulated by the addition of cosolvents. The reported transfer free energies for a model protein from water to alkyl ammonium-based RTILs suggested that all those RTILs act as stabilizers. Yoshimura and coworkers[24] described structural change in chicken egg white lysozyme in the highly concentrated aqueous 1-butyl-3-methylimidazolium nitrate solutions (0-24M) by optical spectroscopy and small-angle X-ray scattering methods. The protein adopted a partially globular state (tertiary structure was disrupted), whereas the protein aggregation was inhibited. The partially globular state was explained by dehydration of the important protein binding sites at high contents of RTIL. Figueiredo and coworkers[25] investigated protein destabilization incurred by 1-butyl-3-methylimidazolium and 1-ethyl-3-methylimidazolium chloride and dicyanamide employing molecular dynamics simulations, NMR spectroscopy and differential scanning calorimetry. According to these authors, stabilization or destabilization of proteins depends on the hydrophilicity of the RTIL anions. Binding of weakly hydrated anions to positively charged or polar residues leads to partial dehydration of the backbone groups and is critical to control protein stability. Thus, dicyanamide is more denaturing than chloride, which should not depend on the cation and, to a reasonable extent, on the concentration.

Amino acid-based ionic liquids (AAILs) represent an interesting and recent implementation of RTILs.[26-31] In AAILs, the anion is obtained from a molecular amino acid via deprotonation, whereas the cation can be picked up in view of prospective usage. Being in tight evolutional connection to alpha amino acids, AAILs are not toxic. They exhibit relatively low ionic conductivities and high shear viscosities, which can be tuned via the addition of water. Miscibility of water and AAILs is excellent over a wide range of contents.[28] Dagade and coworkers[29] reported thermodynamics of ionic hydration and discussed solvent-solute interactions of AAILs in water at room conditions. Woo and coworkers[31] simulated certain equilibrium and transport properties of AAILs. We developed a simple and efficient force field for a significant set of amino acid anions and 1-ethyl-3-methylimidazolium cation.[27]

This work reports a model mini-protein solvation in the three pure AAILs (1-ethyl-3-methylimidazolium alanine, [EMIM][ALA]; 1-ethyl-3-methylimidazolium methionine, [EMIM][MET]; 1-ethyl-3-methylimidazolium tryptophane, [EMIM][TRP]) and their aqueous solutions. Classical empirical-potential molecular dynamics simulation offering an atomistic precision of the results was employed as a primary research technique. The force field of AAILs was developed recently by two of us.[27] We describe structure of the mixed solvent in the vicinity of mini-protein and characterize preferential solvation. We conclude that solvation by AAILs is definitely favorable for the investigated mini-protein, since it better preserves its genuine structure than water.

**Methodology**

Constant-temperature constant pressure molecular dynamics simulations of seven principal systems (Table 1) were conducted at 310 K and 1 bar. Dynamics of the mini-protein and surrounding solvent was recorded during 100 ns. The immediate coordinates of all species were saved every 10 ps, whereas the immediate velocities and forces were not saved. The equations-of-motion were propagated with a time-step of 0.002 ps. The time-step of 0.002 ps requires that the lengths of all intra-molecular (covalent) bonds involving hydrogen atoms were fixed during dynamics. We applied the LINCS algorithm[32] to achieve this. The CHARMM36 force field[33] was used to represent the mini-protein and the recently developed force field was used to represent AAILs.[27] Both force fields are fully compatible. One chloride anion was added to the system to neutralize the mini-protein. Prior to productive runs, the simulated systems were thoroughly equilibrated as described below.

Table 1. List of the simulated systems and their compositions

| # system | AAIL | x (AAIL), mol% | # ion pairs | # water molecules |
|---|---|---|---|---|
| 1 | [EMIM][ALA] | 100 | 600 | 0 |
| 2 | [EMIM][ALA] | 5 | 300 | 5700 |
| 3 | [EMIM][MET] | 100 | 500 | 0 |
| 4 | [EMIM][MET] | 5 | 200 | 3800 |
| 5 | [EMIM][TRP] | 100 | 350 | 0 |
| 6 | [EMIM][TRP] | 5 | 200 | 3800 |
| 7 | water | 0 | 0 | 5065 |

Equilibration of the macromolecule containing systems deserves an extensive attention, since these systems possess lots of local free energy minima. These minima must be avoided ultimately driving the system over a variety of possible states. With the above precautions in mind, we performed the equilibration in the three stages. First, initial geometries were prepared using PACKMOL.[34] We used entry 1L2Y from the Protein Data Bank (PDB) for the TRP-cage mini-protein (Figure 1). Molecular dynamics simulations of 100 ps with no pressure coupling were performed on these geometries using GROMACS 5.0.2 with three-dimensional periodic conditions. The geometry of the mini-protein was restrained at this stage to defend it from any artificially induced conformational changes. Second, the barostat was turned on and the simulations of 10 ns each were conducted at 400 K. The temperature of simulation was temporarily increased to foster equilibration of the AAIL phase. Third, the mini-protein atoms were relieved, and the entire system was simulated for 10 ns at 310 K and 1 bar.

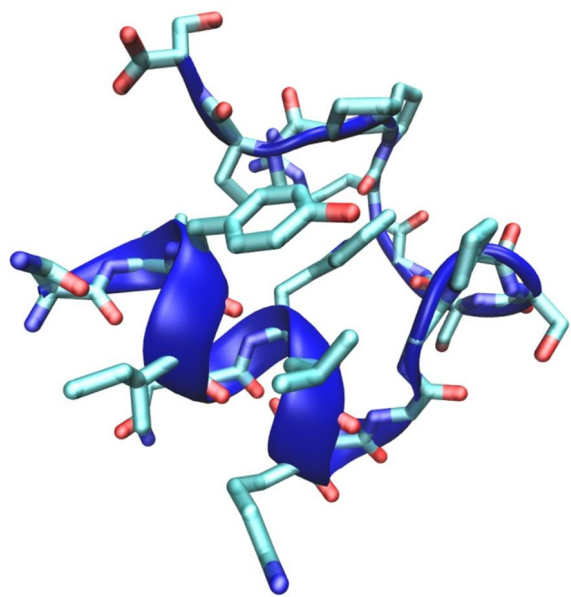

Figure 1. The stick-and-ribbon superimposed representation of the TRP-cage protein (sequence NLYIQ WLKDG GPSSG RPPPS).

The electrostatic interactions in all systems were simulated using direct Coulomb law up to 1.2 nm and using the computationally efficient implementation of the Ewald method (Particle-Mesh-Ewald) beyond 1.2 nm.[35] The Lennard-Jones interactions were smoothly brought down to zero from 1.0 to 1.2 nm using the switched potential approach. The constant temperature was maintained using the Bussi-Donadio-Parinello velocity rescaling thermostat[36] with a relaxation time of 0.1 ps. The constant pressure was maintained using the Parrinello-Rahman barostat[37] with a relaxation time of 1 ps and the compressibility constant of $4.5 \times 10^{-5}$ bar$^{-1}$. The molecular trajectories were analyzed by means of conventional GROMACS utilities and the MDTraj software package[38] using standard definitions of the physical properties of interest. The images have been produced using VMD (Visual Molecular Dynamics) software, version 1.9.[39]

**Results and Discussion**

The molecular configurations of the investigated three-component systems corresponding to free energy minimum at 310 K and 1 bar are depicted in Figure 2. Water and [EMIM][ALA] appear perfectly miscible. Our recent study indicates that AAILs exist as mainly small ionic clusters in water, while water greatly decreases shear viscosity of AAILs.[28] Good mutual miscibility of the simulated cosolvents likely means that both of them are present in the first solvation shell of the mini-protein.

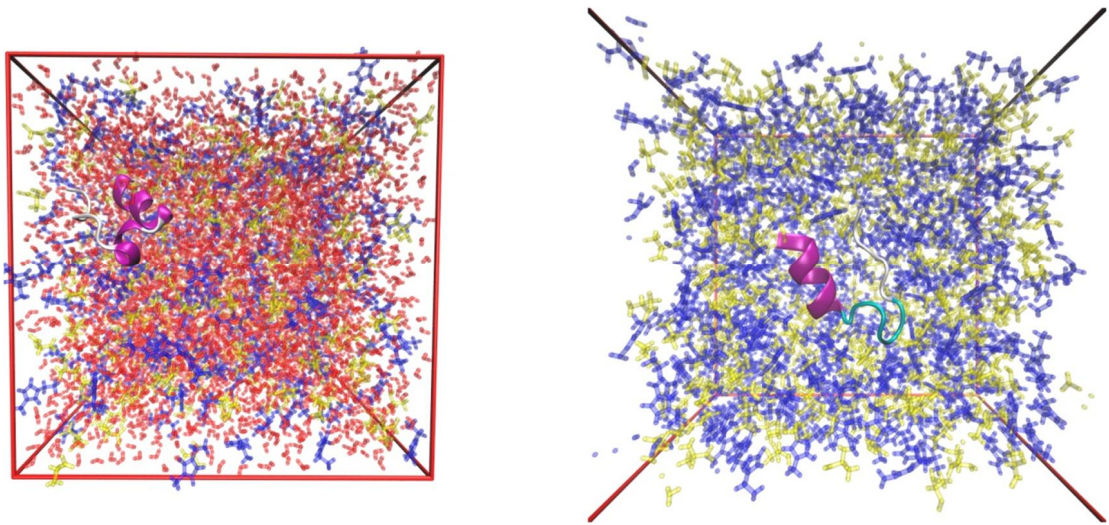

Figure 2. Final molecular configurations of the two selected simulated systems: (left) the mini-protein in system 2 [x([EMIM][ALA]) = 5 mol%]; (right) the mini-protein in system 1 (pure [EMIM][ALA]). The 1-methyl-3-ethylimidazolium cations are blue, the alanine anions are yellow, the water molecules are red. The mini-protein is in cyan and pink.

Figure 3 summarizes radial distribution functions (RDFs) computed between oxygen atoms belonging to water and oxygen atoms belonging to the carboxyl group of the serine residue. The serine residue was chosen because it is polar and, therefore, its strong interaction with polar solvent is expected. Remember that both our cosolvents are polar. Indeed, the first peak is located at 0.28 nm and also the second peak is distinguishable (although small by height) at 0.48 nm. The height of the first peak is smaller in the case of pure water (system 7) than in AAILs (systems 1-6). This trend constitutes a common observation in aqueous solutions. Pure water possesses a very well structured network of hydrogen bonds, which is responsible for high water-water RDF peaks. Interactions with the solute may be strong themselves, but they are always weaker than water-water interactions. That is why local density increase (RDF peak) in the vicinity of protein in pure water is modest. When a significant concentration of another cosolvent is added, it perturbs hydrogen bonds of water, especially upon creation of new hydrogen bonds with water. The resulting water structure in the mixed solvent is weaker than that in pure water. Its binding to the protein produces a higher peak even though the strength of the water-protein interaction remains intact.

The heights of the first peaks in the 5 mol% AAIL solutions are as follows: [EMIM][TRP]>[EMIM][MET]>[EMIM][ALA]. This trend can be correlated to mobility of cosolvents. [EMIM][TRP] is the least mobile AAIL due the anion size, [EMIM][MET] is more mobile, and [EMIM][ALA] is most mobile due to the very small amino acid radical. Coordination number of water in the first coordination sphere of serine amounts to 3.9 (system 7). Coordination numbers of the mini-protein with respect to the anions, $n_-$=3.7 in all AAILs (systems 2, 4, 6). One can conclude that AAIL substitutes most water molecules in the vicinity of the mini-protein in the 5 mol% AAIL solutions.

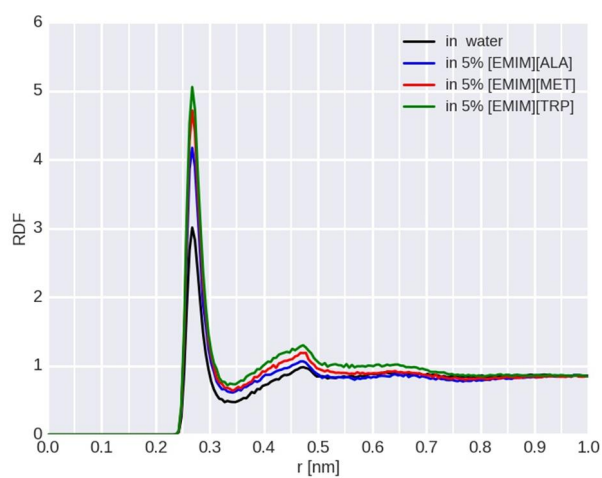

Figure 3. Radial distribution functions (RDF) featuring spatial correlation between carboxyl oxygen atoms of the serine residue (mini-protein) and oxygen atoms (water) in systems 2, 4, and 6.

We selected an intrinsically acidic hydrogen atom of the imidazole ring to represent binding behavior of the 1-ethyl-3-methylimidazolium cation. This site of the cation was selected, because it is most electron deficient among atoms of [EMIM]$^+$. Particularly this atom participates in hydrogen bonding with other chemical entities including anions in the condensed phase of ionic liquid. Figure 4 depicts RDFs between serine residue (carboxyl oxygen atom) and imidazole hydrogen atom computed in the six simulated AAIL containing systems. The amino acid anion of AAIL impacts significantly and provides different RDFs in each AAIL.

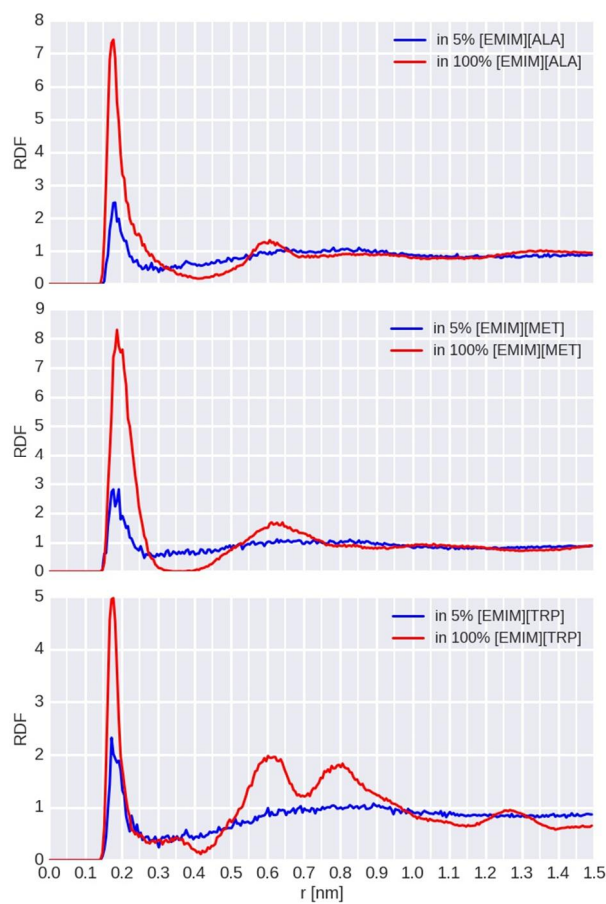

Figure 4. Radial distribution functions (RDF) featuring spatial correlation between carboxyl oxygen atoms of serine residue (mini-protein) and intrinsically acidic hydrogen (imidazole ring) in systems 1, 2, 3, 4, 5, and 6.

The first peaks in all AAIL containing systems is located at 0.18 nm, which corresponds to a strong hydrogen bond between the imidazolium-based cation and the mini-protein. The first minima are poorly defined ranging from 0.3 to 0.4 nm. As composition of the first coordination sphere directly depends on the definition of it, the corresponding coordination numbers with respect to 1-ethyl-3-methylimidazolium cation, $n_+$, must be treated with caution. In the 5% AAIL solutions, $n_+$=0.11 in [EMIM][ALA], $n_+$=0.11 in [EMIM][MET] and

$n_+$=0.07 in [EMIM][TRP]. Compare with $n_-$ reported above. The mutual affinity between the mini-protein and amino acid anions are hereby proven.

In the pure AAIL systems (1, 3, 5), $n_+$ obviously increases: $n_+$ ([EMIM[ALA])=0.92, $n_+$ ([EMIM[ALA])=0.77, $n_+$ ([EMIM[ALA])=0.23. This is in concordance with RDFs for the mini-protein and the anions suggesting that affinity of the anions decreases in the following row: [EMIM][TRP]>[EMIM][MET]>[EMIM][ALA].

The root-mean-square deviation (RMSD) of the macromolecule structure is a central measure of its stability, in the broad sense, vs. time. Rupture of the initial mini-protein structure would result in high values of RMSD. In turn, smaller RMSD values would suggest a better conservation of the protein. RMSD is not only the function of solvent transport properties, but also – and to a larger extent – on specific chemical interactions between the mini-protein and solvents. According to Figure 5, the mini-protein has the largest conformational flexibility in water. Water is a natural environment for proteins. Although the investigated mini-protein retains stability at 310 K, temperature elevation will likely violate an initial structure drastically. It is widely known that many proteins in the multicellular organisms start denaturing at 40-50 °C. Such a denaturation is a primary reason that very little higher animal are able to survive even short temperature elevations. Bacteria and virus are much more advanced in this aspect, because their proteins are different.

Thermal denaturation is considered the most frequent denaturation pathway, which takes place during storage of macromolecules in vitro. The conventional techniques, such as chemical modification, immobilization, stabilizing agents, cannot defend from thermal denaturation. However, this can be realistically achieved through the usage of different solvent, because RMSD in pure AAILs is significantly smaller than in water and in 5 mol% aqueous mixtures of AAILs. The difference between the cases of pure [EMIM][ALA], [EMIM][MET], and [EMIM][TRP] is not drastic. All three AAILs efficiently solvate the mini-protein similarly to water. They do not induce any observable structural alterations. Moreover, thermal fluctuations of the mini-protein structure in AAILs are smaller suggesting a better thermal stability of the mini-protein.

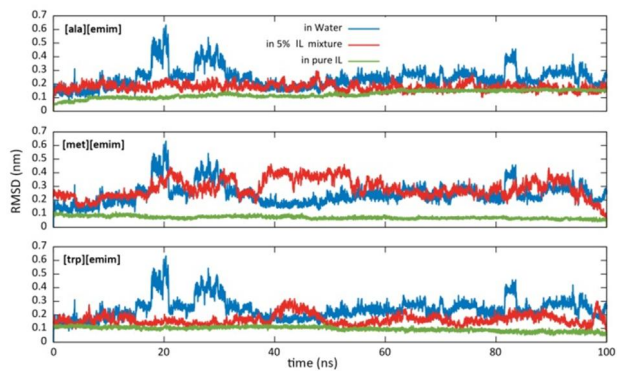

Figure 5. Evolution of the root-mean-square deviation (RMSD) of the mini-protein structure upon the 100 ns long equilibrium molecular dynamics simulations. The first equilibrated geometry was taken as a reference point. See legends for system designation.

Radius of gyration ($R_g$, Figure 6) confirms conclusions about a higher stability of the mini-protein in AAILs and 5 mol% AAIL solutions in water. Little changes of $R_g$ were observed during 100 ns of equilibrium molecular dynamics simulations. $R_g$ of the mini-protein in [EMIM][ALA] equals to 0.74±0.01 nm; 0.76±0.01 nm in [EMIM][MET]; 0.74±0.01 nm in [EMIM][TRP]; 0.74±0.01 nm in pure water. The standard deviation of $R_g$ in water is slightly larger than in other systems, which is consistent with larger RMSD (Figure 5). All corresponding values of $R_g$ and their components are summarized in Table S1.

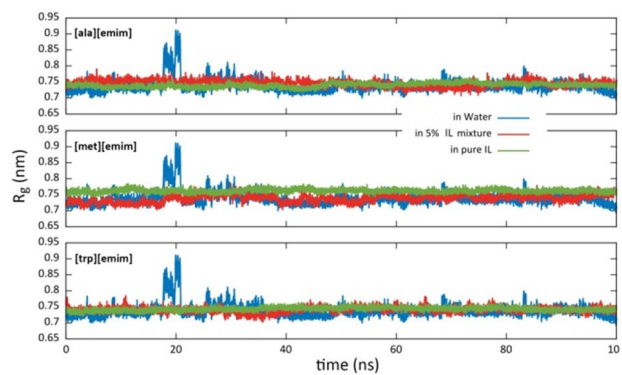

Figure 6. Radius of gyration of the mini-protein as a function of time during equilibrium molecular dynamics simulations. See legends for system designation. The average values of radius of gyration are summarized in Table S1.

One of the most interesting questions that must be attended in this investigation is which solvent and which chemical entity in particular prevail in the first coordination sphere of the model mini-protein and are, therefore, responsible for its efficient solvation and conservation. The posed question can be answered in terms of distribution functions and coordination numbers, which are the integrals of those functions. Figures 7-8 investigate spatial correlations between terminal hydrogen atoms (position $\omega$) and methyl group hydrogen atoms of the mini-protein with oxygen atoms of water and three AAILs during the 100 ns long equilibrium molecular dynamics simulations of the seven systems. The two most representative plots are shown here. An interested reader is referred to Supplementary Information, Figures S1-S4, for all summarized RDFs. The $\omega$-H atoms are strongly correlated with carboxyl groups of the anions by creating multiple hydrogen bonds with the length of ca. 0.18 nm. In turn, both imidazolium cations and water molecules are pushed to the second coordination sphere, where they exhibit quite smashed RDF peaks. Unlike [EMIM]$^+$, certain amount of water molecules is still present in the first shell, but the corresponding RDF maximum is much smaller as compared to that of the anions.

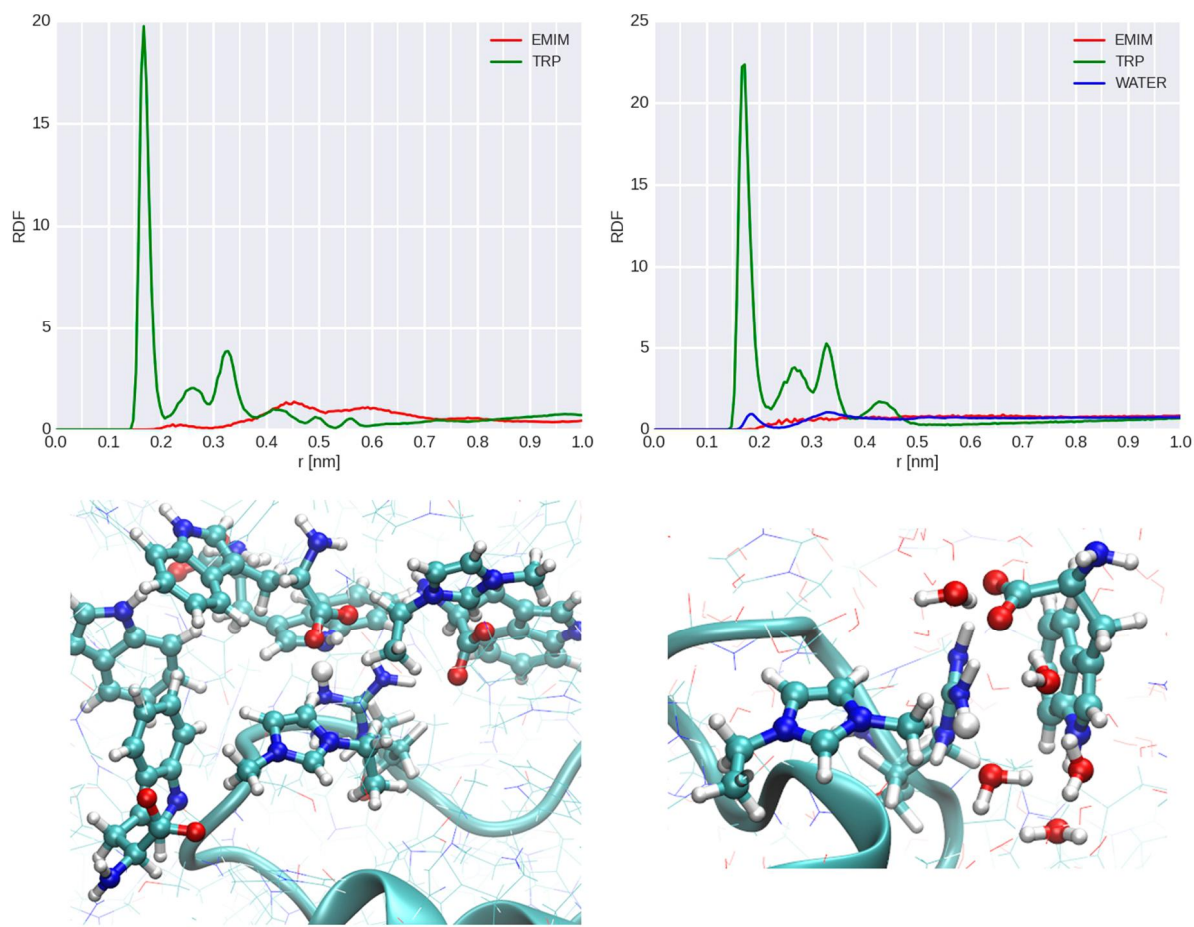

Figure 7. Radial distribution functions (RDFs) featuring spatial correlation between hydrogen atoms (position ω) of the mini-protein and oxygen atoms of solvents in pure AAIL (left) and 5% [EMIM][TRP] AAIL/water mixture (right). Snapshots highlight mini-protein solvation.

Binding of all solvents to the methyl group is sufficiently different. Despite the first peak corresponding to [EMIM]$^+$ at 0.42 nm, no strong structure correlations were observed. Even this peak vanishes in the 5 mol% [EMIM][MET] aqueous solution. The RDFs in other AAILs (Figures S1-S2) completely support this conclusion. All applicable coordination numbers range between 0.2 and 0.3. That is the methyl group appears insufficiently solvated by both solvents. Based on the RDF in the pure [EMIM][TRP] system, one can identify that 310 K is lower than a glassy transition temperature for the system containing the model mini-protein and this AAIL (system 5).

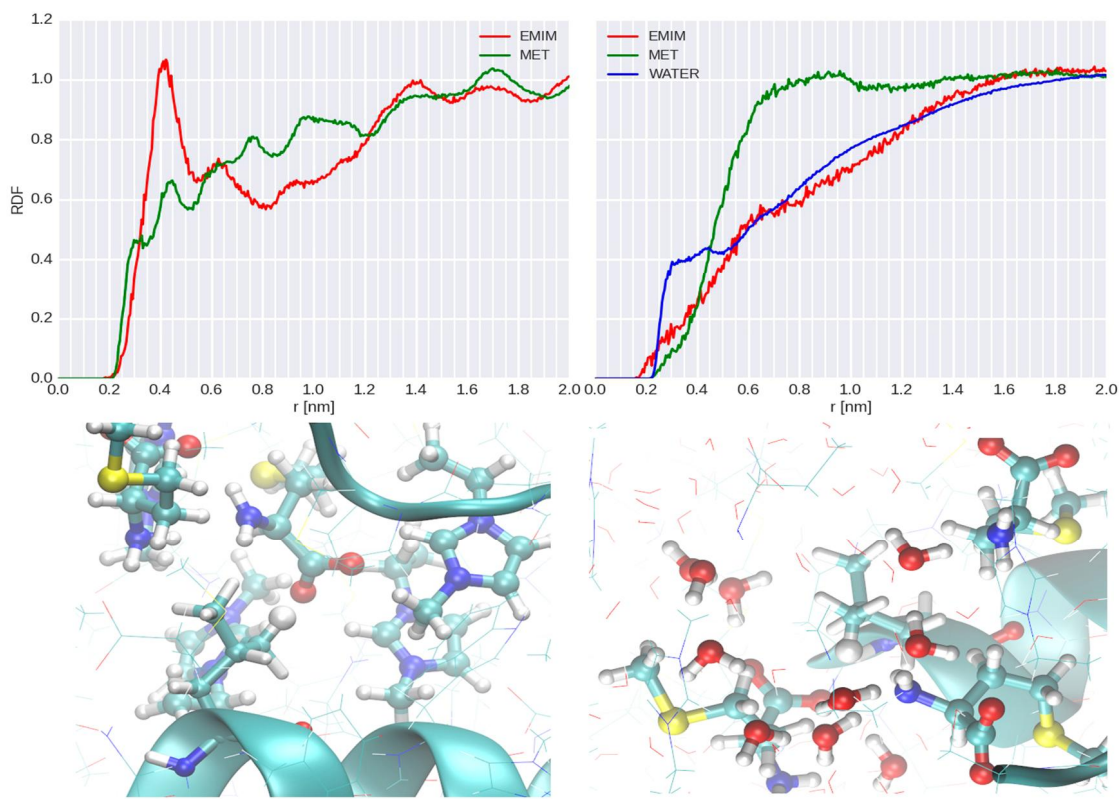

Figure 8. Radial distribution functions (RDFs) featuring spatial correlation between hydrogen atoms (methyl groups) of the mini-protein and oxygen atoms of solvents in pure AAIL (left) and 5% [EMIM][MET] AAIL/water mixture (right). Snapshots highlight mini-protein solvation.

**Conclusions**

Classical molecular dynamics simulations using the recently developed force field for amino acid-based ionic liquids were carried out to observe their solvation behavior in relation to model mini-protein. In addition to pure AAILs, 5 mol% aqueous solutions of these AAILs were considered, since water is always abundant in real biological systems. We found that AAILs efficiently coordinate the mini-protein. This happens primarily thanks to amino acid based anions, which interact strongly with (the 'like dissolves like' rule holds). AAILs push some water molecules outside the first coordination sphere. This observation is in concordance with previous reports.[7,10] Being coordinated by the ions, the mini-protein exhibits hindered mobility, as indicated by the RMSD analysis in all six AAIL containing systems. The conformation flexibility is even smaller when water is completely excluded. This observation favors applications of amino acid-based ionic liquids for conservation of proteins and their subsequent usage in vitro. Importantly, solvation by AAILs does not perturb an initial shape and configuration of the mini-protein, as suggested by constant radius of gyration and its components. According to our computational analysis, AAILs emerge as an interesting candidate solvent or cosolvent for a robust protein solvation, conservation and storage. We anticipate that our paper will inspire more experimental efforts in this direction.

## Acknowledgments

The research was supported by CAPES, FAPESP and CNPq.

## Supporting Information

Table S1 summarizes radii of gyration and their components. Figures S1-S4 show radial distribution functions for the mini-protein and selected AAILs.

## Author Information

E-mails for correspondence: chevrot@sdu.dk (G.C.); fileti@gmail.com (E.E.F.); vvchaban@gmail.com (V.V.C.)